\documentclass[a4paper]{article}
\usepackage{graphicx}
\begin{document}
\title{Renormalized perturbation calculations for the single impurity Anderson model.}
\author{A.C.Hewson\\  Dept. of Mathematics, Imperial
College, London SW7 2BZ.\\}
\date{}
\maketitle

\begin{abstract}
We illustrate the renormalized perturbation expansion method  by applying it to a single impurity Anderson model.
Previously, we have shown  that this approach gives the {\it exact}
leading order results for the specific heat, spin and charge susceptibilities and leading order
temperature dependence of the resistivity for this model in the Fermi-liquid regime,  when carried out to second order in the renormalized interaction $\tilde U$. Here we consider the effects of higher order quasi-particle scattering and calculate the third order contributions to the
$H^3$ term in the impurity magnetization for the symmetric model in a weak magnetic field $H$. The result is asymptotically exact in the weak coupling regime, and is very close to the exact Bethe ansatz result in the Kondo regime. We also calculate the quasi-particle density of states
in a magnetic field, which is of interest in relation to recent experimental work on quantum dots.
 \end{abstract}

{\centerline{ email address: a.hewson@ic.ac.uk}\par
\pagestyle{empty}
\section{\bf Introduction}\par
Systems with strong local inter-electron interactions
have been the focus of much theoretical work in recent years, as these include a variety
of interesting systems
ranging from high $T_c$ superconductors, heavy fermions and Mott insulators, to mesoscopic
systems such as quantum dots. Conventional perturbation theory cannot deal with
strong interactions in general, so that new techniques have to be developed to make predictions
for the behaviour of such systems. Specialized techniques, such as the Bethe ansatz or
conformal field theory,  have been successfully developed for certain
classes of systems, such as one dimensional systems and impurity models. However, techniques are
required that can be applied more generally, particularly for systems in two and three dimensions.
One approach which has been extended to a wider class of problems is the numerical renormalization
group approach (NRG) as developed by Wilson \cite{wil},  which was orginally successfully applied 
to models of magnetic impurities. In this approach the higher energy excitations are progressively 
 eliminated to deduce a sequence of effective models for the behaviour on lower and lower energy scales. The behaviour on the lowest energy scales can be calculated from the limiting fixed point Hamiltonian of this sequence, and its leading correction terms. In its original form the method
only works for impurity systems but a modified form of the approach, the density matrix renormalization group method (DMRG) \cite{whi,dmrg},  has been  successfully developed for one  dimensional systems. In principle the DMNRG could  be extended to two and three dimensions  but there are technical difficulties in practice, though some
calculations for two dimensional systems have been carried out. In another development the NRG has also been extended to higher dimensional lattice models by use of  dynamical mean field theory (DMFT) \cite{dmft}. This approach exploits the fact that certain infinite dimensional lattice models, such as the Hubbard and periodic Anderson models, can  mapped onto effective impurity models, together with a self-consistency condition \cite{ss,bul}.  The calculations for the effective impurity models can  be carried out using the NRG and  iterated until the self-consistency
condition is satisfied. The DMFT has considerably extended the potential range of application of the NRG approach.  The NRG, however,
is not the only way of realizing  renormalization group ideas. The earlier way of applying the renormalization group, as originally developed in field theory,
was via a reorganisation of the perturbation expansion, such that the  expansion could be carried out
in terms of the renormalized parameters. This rearrangement of perturbation theory enabled one to circumvent the problem of  the divergences which
had plagued the conventional  approaches. The elimination of the divergences, however,
was essentially a by-product of this approach and it is possible to use the reorganisation of the perturbation expansion as a strategy for
dealing with the low energy behaviour of a wide variety  systems. The renormalized perturbation theory approach could be particularly useful in situations where there are strong renormalizations of the
basic parameters, such as in the Fermi-liquid regime for heavy fermions,
 where the masses of the electrons may be renormalized by a factors of the order of 1000.\par
In earlier work \cite{h-rpt} we have shown how the renormalized perturbation theory can be applied to impurity models. In particular, we have shown that this approach, when applied to the Anderson impurity model and carried out to second order in the renormalized interaction $\tilde U$,  gives the exact
leading order results for the specific heat, spin and charge susceptibilities at $T=0$, and the leading order
temperature dependence of the resistivity in the Fermi-liquid regime. In this paper we begin with a brief review of these earlier results, and  we then calculate  some higher order corrections, in particular, we calculate the $H^3$ term  for the impurity magnetization
in a weak magnetic field $H$  to third order in the renormalized interaction term $\tilde U$, and compare the results with the exact Bethe ansatz result. In the final section we calculate the quasiparticle density of states  in
a magnetic field. These latter results are of some interest in comparing with
the low temperature linear response results on quantum dots in the presence of a magnetic field. 
\section{\bf Renormalized Perturbation Theory}\par
We formulate the renormalized perturbation expansion for the impurity Anderson model \cite{and}. 
 The Hamiltonian
for this model is
\begin{equation} H=\sum\sb {\sigma}\epsilon\sb {d}
d\sp {\dagger}\sb {\sigma}
d\sp {}\sb {\sigma}+
Un\sb {d,\uparrow}n\sb {d,\downarrow}
 +\sum\sb {{ k},\sigma}( V\sb { k}d\sp {\dagger}\sb {\sigma}
c\sp {}\sb {{ k},\sigma}+ V\sb { k}\sp *c\sp {\dagger}\sb {{
k},\sigma}d\sp {}\sb {\sigma})+\sum\sb {{
k},\sigma}\epsilon\sb {{ k},\sigma}c\sp {\dagger}\sb {{ k},\sigma}
c\sp {}\sb {{
k},\sigma},\label{ham}\end{equation}
which describes an impurity d level $\epsilon_d$, 
  hybridized with 
 conduction electrons of the host metal via a matrix element $V_k$, with an
 interaction term $U$ between the electrons in the localized d state, where $n_{d,\sigma}=d^{\dagger}_{\sigma}d^{}_{\sigma}$.
The width of the localized bound state for $U=0$ depends on
the quantity  $ \Delta(\omega)=\pi\sum\sb {k}|
 V\sb {k}|\sp 2\delta(\omega -\epsilon\sb {
k})$. For a  conduction without any prominant features this function
does not have a strong dependence on $\omega$, so it is usual to
 take the case of a  wide  conduction band  with a flat  density of states 
 where $\Delta(\omega)$ 
becomes independent of $\omega$ and can be taken as a constant $\Delta$.
The partition function $Z$ for this model can be expressed as a functional integral
over imaginary time dependent Grassmann variables, corresponding to the electron creation and
annihilation operators, using the standard methods (see for example \cite{no}),

\begin{equation}Z=\int \prod_{\sigma}{\cal D}(\bar d_{\sigma}){\cal D}(d_{\sigma}){\cal D}(\bar c_{k,\sigma}){\cal D}(c_{k,\sigma})e^{-S},\end{equation}
where the action $S$ is given by
\begin{equation}S=\int_0^{\beta} {\cal L}_{\rm AM}(\tau)d\tau, \end{equation}
and the Lagrangian for the Anderson model is given by
$${\cal L}_{\rm AM}(\tau)=
\sum_{\sigma}\bar d_{\sigma}(\tau)(\partial_{\tau}-\epsilon_d)d_{\sigma}(\tau)+\sum_{k,\sigma}
c_{k,\sigma}(\tau)(\partial_{\tau}-\epsilon_k)c_{k,\sigma}(\tau)$$
\begin{equation}+Un_{\uparrow}(\tau)n_{\downarrow}(\tau)+\sum_{\sigma}V_k(\bar d_{\sigma}(\tau)c_{k,\sigma}(\tau)+\bar c_{k,\sigma}(\tau)d_{\sigma}(\tau)),\end{equation}
where
$n_{\sigma}(\tau)=\bar d_{\sigma}(\tau)d_{\sigma}(\tau)$. One can formally integrate over the Grassmann variables for the conduction electrons, as they involve only quadratic terms,
and express the result in terms of a reduced action $S_{\rm red}$, 
\begin{equation}Z=\int \prod_{\sigma}{\cal D}(\bar d_{\sigma}){\cal D}(d_{\sigma})e^{-S_{\rm red}},\end{equation}
where  $S_{\rm red}$ is given by
\begin{equation}S_{\rm red}=\int_0^{\beta}d\tau\int_0^{\beta}d\tau'  \sum_{\sigma}\bar d_{\sigma}(\tau)[G_{\sigma}^{(0)}(\tau-\tau')]^{-1}d_{\sigma}(\tau')+U\int_0^{\beta}d\tau n_{\uparrow}(\tau)n_{\downarrow}(\tau),\end{equation}
with $G_{\sigma}^{(0)}(\tau)=1/\beta\sum_nG_{\sigma}^{(0)}(i\omega_n)e^{-i\omega_n\tau}$, where $\omega_n=(2n+1)\pi/\beta$.  The non-interacting Green's function for the localized
electron $G_{\sigma}^{(0)}(i\omega_n)$ given by
\begin{equation}G_{\sigma}^{(0)}(i\omega_n)={1\over {i\omega_n+\sigma h-\epsilon_d +i\Delta{\rm sgn}(\omega_n)}},\end{equation}
where we have included a coupling to a magnetic field $H$, and $h$ given by  $h=g\mu_{\rm B} H/2$.
The Grassmann variables are required to satisfy antiperiodic boundary conditions $\bar d(\beta)=-\bar d(0)$ and $d(\beta)=- d(0)$.\par
The Fourier transform of the corresponding  retarded
 one-particle double-time Green's function $G^{(0)}_{{\rm d}\sigma}(\omega)$
for the localized d electron can be deduced by analytical continuing to real frequencies
$i\omega_n\to\omega-i\delta$ ($\delta\to +0$). On introducing a corresponding self-energy $\Sigma_{\sigma}(\omega,h)$ the interacting
retarded Green's function can be written in the form,
\begin{equation}G_{\sigma}(\omega)=
{1\over {\omega-\epsilon_{\rm d}+\sigma h
+
i\Delta }-\Sigma_{\sigma}(\omega,h)}.\label{gf}\end{equation} 
In the conventional perturbation expansion this self-energy is calculated in powers of the 
interaction $U$. It will be convenient to write the self-energy in the 
form $\Sigma_{\sigma}(\omega+\sigma h,h)$ because the non-interacting Green's functions, which are the propagators in the perturbation expansion, are functions of the combined variable $\omega+\sigma h$. In the renormalized perturbation theory the perturbation  expansion is reorganized
to a form appropriate for the low energy 
regime.  The  first step is to write the self-energy in the form,
\begin{equation}\Sigma_{\sigma}(\omega+\sigma h,h)=\Sigma_{\sigma}(0,0)
+(\omega+\sigma h)\Sigma'_{\sigma}
(0,0)+\Sigma^{\rm rem}_{\sigma}(\omega,h),\label{self}\end{equation}
which does nothing more than define  the remainder
 self-energy $\Sigma^{\rm rem}_{\sigma}(\omega+\sigma h,h)$, except that we have assumed that  
Luttinger's result \cite{lut} that  $\Sigma'_{\sigma}(0,0)$ is real. When this is substituted
back into equation (\ref{gf}), the Green's function takes the same form with a
 `renormalized'
 energy level, width of the localized state and self-energy, which are denoted by  a tilde, defined by
\begin{equation}\tilde\epsilon_{\rm d}=z(\epsilon_{\rm d}
+\Sigma_{\sigma}(0,0)),\quad\tilde\Delta =z\Delta ,\quad \tilde\Sigma_{\sigma}(\omega,h)=z\Sigma^{\rm rem}_{\sigma}(\omega,h),\label{ren1}\end{equation}
where $z$,  the wavefunction renormalization
 factor, is given by
$z={1/{(1-\Sigma'_{\sigma}(0,0))}}$, and $\Sigma_{\sigma}(0,0)$ and $\Sigma'_{\sigma}(0,0)$
are to be evaluated at $T=0$ as well as $\omega=h=0$. These will be the parameters of the renormalized theory
instead  of $\epsilon_{\rm d}$ and $\Delta$,
which are specified in the 'bare' Hamiltonian of equation (\ref{ham}). Note that
the g-factor coupling to the magnetic field $H$ is unrenormalized.
The overall $z$-factor is removed by rescaling the Grassmann fields, $d_{\sigma}(\tau)\to \sqrt{z}\tilde
d_{\sigma}(\tau)$. \par
\smallskip\noindent 
 The last parameter specifying the renormalized theory is the renormalized
interaction $\tilde U$.  This quantity is derived from the irreducible four point
vertex function  $\Gamma_{\sigma,\sigma'}(\omega,\omega')$,
 which is a special case of the more general irreducible four point
vertex function $\Gamma^{\sigma'',\sigma'''}_{\sigma,\sigma'}
(\omega,\omega';\omega'',\omega''')$ with $\sigma''=\sigma$, 
$\sigma'''=\sigma'$, $\omega''=\omega$ and $\omega'''=\omega'$.
This latter quantity is derived from the two particle Green's function
of the d-electrons in the usual way. The renormalized four
 point vertex function is defined by $\tilde\Gamma_{\sigma,\sigma'}
(\omega,\omega')=z^2\Gamma_{\sigma,\sigma'}(\omega,\omega')$,
and takes account of the rescaling of the local fermion fields. 
 The renormalized interaction $\tilde U$ is then defined by the value
 of $\tilde\Gamma_{\sigma,\sigma'}(\omega,\omega')$ at 
 $\omega=\omega'=0$,  
\begin{equation}\tilde U=\tilde\Gamma_{\sigma,\sigma'}(0,0).\label{ren2}\end{equation}\par
As certain of the interaction effects are taken into account {\it ab initio}
in the renormalized theory compensating terms have to be introduced to
avoid overcounting. The Lagrangian for Anderson model can be rewritten in the form, \begin{equation}{\cal L}_{\rm AM}(\bar d_{\sigma},d_{\sigma},\epsilon_d,\Delta,U)={\cal L}_{\rm AM}(\tilde{\bar d_{\sigma}},\tilde d_{\sigma},\tilde\epsilon_d,\tilde\Delta,\tilde U)+{\cal L}_{\rm CT}(\tilde{\bar d_{\sigma}},\tilde d_{\sigma},\lambda_1,\lambda_2,\lambda_3),\end{equation}
in terms of the renormalized fields, where the counter-term Lagrangian is given by 
\begin{equation}{\cal L}_{\rm CT}(\tilde{\bar d_{\sigma}},\tilde d_{\sigma},\lambda_1,\lambda_2,\lambda_3)=\tilde {\bar d_{\sigma}}(\tau)(\lambda_2\partial_{\tau}
+\lambda_1)\tilde d_{\sigma}+\lambda_3\tilde n_{\uparrow}(\tau)\tilde n_{\downarrow}(\tau)\end{equation}
where $\lambda_1=-z\Sigma(0,0)$, $\lambda_2=z-1$  and $\lambda_3=z^2(U-\Gamma_{\uparrow,\downarrow}(0,0))$.
By construction the renormalized self-energy  
$\tilde\Sigma_{\sigma}(\omega)$ is such that
\begin{equation}\tilde\Sigma_{\sigma}(0,0)=0,\quad \tilde\Sigma'_{\sigma}
(0,0)=0,\label{r1}\end{equation}
\smallskip\noindent so that $\tilde\Sigma_{\sigma}(\omega)
 =O(\omega^2)$
 for small $\omega$,  on the assumption that it is analytic
 at $\omega=0$.
 As $\tilde\Gamma_{\sigma,\sigma}(0,0)=0$  we also have
\begin{equation}\tilde\Gamma_{\sigma,\sigma'}(0,0)=\tilde U
(1-\delta_{\sigma,\sigma'})
.\label{gamma}\label{r2}\end{equation}
\par
The quasiparticle or renormalized Green's function 
 takes the form 
\begin{equation}\tilde G_{\sigma}(\omega)=
{1\over {\omega-\tilde\epsilon_{\rm d}+\sigma h
+i\tilde\Delta -\tilde\Sigma_{\sigma}(\omega,h)}}.
\label{qpgf}\end{equation}
The reorganized  perturbation theory is set up to calculate the renormalized self-energy
 $\tilde\Sigma_{\sigma}(\omega,h)$.  
 The propagators in this expansion correspond to the non-interacting quasiparticles in the Lagrangian ${\cal L}_{\rm AM}(\bar d_{\sigma},d_{\sigma},\epsilon_d,\Delta,U)$ with $\tilde U=0$. The quasiparticle interaction $\tilde U$ is used as
an expansion parameter but all the terms in  the counter Lagrangian
${\cal L}_{\rm CT}$
have to be included as well. To organize the expansion in powers of $\tilde U$
the terms $\lambda_1$, $\lambda_2$
 and $\lambda_3$ have also to be  expressed formally as a powers in $\tilde U$, 
\begin{equation}\lambda_{1}=\sum_{n=0}^{\infty}\lambda_1^{(n)}\left({\tilde U\over\pi\tilde\Delta}\right)^n,\quad
\lambda_{2}=\sum_{n=0}^{\infty}\lambda_2^{(n)}\left({\tilde U\over\pi\tilde\Delta}\right)^n,\quad 
\lambda_3=\sum_{n=0}^{\infty}\lambda_3^{(n)}\left({\tilde U\over\pi\tilde\Delta}\right)^n.\label{ps}\end{equation}
 The coefficients $\lambda_1^{(n)}$,
 $\lambda_2^{(n)}$ and $\lambda_3^{(n)}$ are then determined by the requirement
 that the three normalization conditions, (\ref{r1}) and (\ref{r2}), are satisfied to each order 
in the 
expansion. These normalization conditions are essentially those used within field
theory in order to circumvent the problem of infinities arising from the lack of an ultraviolet cut-off (see for example \cite{ft}). The procedure, however, makes no mention of infinities, it simply  allows the field theoretic perturbation expansion to be expressed in terms of the experimentally observed masses and interactions.
In condensed matter systems divergences do not arise in this way, as there is always a natural cut-off, so this is no necessity to reorganise the perturbation expansion. However, for
the Anderson model there are very strong renormalizations of the effective d-level and the interactions at low energies in the Kondo regime, where the impurity
d-electrons are virtually localized, which make it desirable when working in this regime to
take account of these very strong renormalizations from the start. It allows one to develop a
perturbation theory with an effective d-level and interactions appropriate to this energy scale. 
More generally it makes a direct link to Landau Fermi liquid theory. There are no cut-off dependent ultraviolet divergences to eliminate so the question of the renormalizability of the model does not arise.
\par 
We can obtain significant results with this approach even at zero order, $\tilde U=0$. If we calculate the
quasiparticle occupation number $\tilde n_{{\rm d},\sigma}$ at $T=0$ and $H=0$ from (\ref{qpgf}) with
$\tilde U=0$, we find
\begin{equation}\tilde n_{{\rm d},\sigma}={1\over
2}-{1\over\pi}\tan ^{-1}\left
({\tilde\epsilon_{{\rm d}}}\over{\tilde\Delta}\right
).\label{qpfsr}\end{equation}
As the factors of z in this expression cancel it is equivalent to the exact Friedel sum rule 
\cite{fsr} and
 expresses the one-to-one correspondence between the quasiparticle number and electron number
in Landau Fermi liquid theory \cite{lan}, $\tilde n_{{\rm d},\sigma} = n_{{\rm d},\sigma}$. Hence, the d-level occupation at $T=0$ can be calculated 
from the zero order renormalized Green's function. \par
\noindent The Freidel sum rule  also holds in the presence of a magnetic
field and the equivalent expression for  the occupation of the d-level in
terms of the renormalized self-energy is  given by
\begin{equation} n_{{\rm d},\sigma}={1\over
2}-{1\over\pi}\tan ^{-1}\left
({\tilde\epsilon_{{\rm d}}-\sigma h+\tilde\Sigma_{\sigma}(0,h)}\over{\tilde\Delta}
\right
).\label{rfsr}\end{equation}

We have to use the perturbation theory to calculate the field dependence of the
renormalized self-energy. However, we can show that it is sufficient  to work only to first order 
in $\tilde U$ to obtain the exact result. There are two terms to first order,
one from the 
 tadpole or Hartree diagram,  and one from the corresponding counter-term diagram in $\lambda_1$. There is no wavefunction renormalization to this order
 so $\lambda_2^{(1)}=0$ and also to this order $\tilde\Gamma_{\uparrow,\downarrow}(0,0)=\tilde U$, so $\lambda^{(1)}_3=0$. To satisfy the renormalization conditions the counter-term should cancel the contribution
from the tadpole diagram for $T=h=0$ so $\lambda_1=
\tilde Un^{(0)}_{d,-\sigma}(0,0)$. 
Hence the combined contribution is
\begin{equation}\tilde\Sigma_{\sigma}^{(1)}(\omega, h,T)=\tilde U
(n^{(0)}_{d,-\sigma}(h,T)-
n^{(0)}_{d,-\sigma}(0,0)).\label{first}\end{equation}
\smallskip 
The spin susceptibility of the d-electrons at $T=0$ 
can be calculated  from $g\mu_{\rm B}(n_{{\rm d},\uparrow}-n_{{\rm d},\downarrow})/2$, 
 by substituting the self-energy from 
(\ref{first}) into  equation (\ref{rfsr}), and then differentiating with respect to $H$.
The  charge susceptibility can be calculated in a similar way and the two results are
 \begin{equation}\chi_{d}={{(g\mu_{\rm B})^2}\over 2}\tilde\rho_{d}(0)(1+
 \tilde U\tilde\rho_d(0)),\quad\chi_{d,{\rm c}}=2\tilde
 \rho_d(0)(1-\tilde
 U\tilde\rho_d(0)),\label{rsus}\end{equation}
where $\tilde\rho_d(0)$ is the quasiparticle density of states at the Fermi level and is given by
\begin{equation}\tilde\rho_d(0)={\tilde\Delta/\pi\over{\tilde\epsilon_d^2+\tilde\Delta^2}}\label{qpdos}.\end{equation}
 It is not obvious that these results to first order in $\tilde U$ are exact.
However, there are  Ward 
identities \cite{yam} which can be derived from charge and spin conservation, and
in terms of the renormalized self-energy and density of states take the form,
 \begin{equation}{{\partial{ \tilde\Sigma}_{\sigma}(\omega)}\over
   {\partial
  h}}\biggr |_{\omega=0}={{\partial{ \tilde\Sigma}_{\sigma}(\omega)}
  \over {\partial
  \mu}}\biggr |_{\omega=0}=-\tilde\rho_{{\rm d},\sigma}(0)\tilde
  U.\label{wi}\end{equation}
The spin and charge susceptibilities can be  derived from
these exact relations on using (\ref{rfsr}) to give
 \begin{equation}\chi_{d}={{(g\mu_{\rm B})^2}\over 2}\tilde\rho_{d}(0)(1-\partial\tilde
 \Sigma/\partial h)={{(g\mu_{\rm B})^2}\over 2}\tilde\rho_{d}(0)(1+
 \tilde U\tilde\rho_d(0)),\end{equation} 
 and
   \begin{equation}\chi_{d, {\rm c}}=2\tilde\rho_d(0)(1+\partial\tilde\Sigma/\partial\mu)=2\tilde
 \rho_d(0)(1-\tilde
 U\tilde\rho_d(0)),\end{equation}
confirming that the first order results for these quantities are indeed exact. \par
The impurity contribution to the low temperature specific heat coefficient from the
non-interacting quasiparticles ($\tilde U=0$) is given simply by
\begin{equation}\gamma_{d}={{2\pi^2}\over
3}\tilde\rho_{\rm d}(0).\label{gamma}\end{equation}
This result corresponds to the exact result calculated by Yamada \cite{yam}.
 It is a general feature of Fermi liquid theory that the quasiparticle
interactions do not give any corrections to the linear coefficient of specific heat 
as their contributions to the specific heat are of higher order in temperature.\par
In the local moment or Kondo regime the local charge susceptibility must
go to zero, so from equation (\ref{rsus}) we find $\tilde U\tilde\rho_d(0)=1$.
If we define the Kondo temperature $T_{\rm K}$ by $\chi_d=(g\mu_{\rm B})^2/4 T_{\rm K}$  then $\tilde\rho_d(0)=1/4 T_{\rm K}$  and all the results can
be written in terms of $T_{\rm K}$.  They correspond to the exact results
for the s-d or Kondo model \cite{ba,tw}.\par
From the exact Bethe ansatz results \cite{ba,tw} for the spin and charge susceptibility for the symmetric Anderson model 
it is possible to deduce the renormalized parameters, $\tilde\Delta$ and $\tilde U$, in
terms of the bare parameters $\Delta$ and $U$. These are are shown in figure 1. Initially
$\tilde U\sim U$ for small $U$, but when $U/\pi\Delta>2$, the energy scales $\tilde U$ and 
$\pi\tilde\Delta$ merge in the strong coupling regime and $\tilde U=\pi\tilde\Delta=4T_{\rm K}$.  \par
\begin{figure}
\begin{center}
\includegraphics[width=10cm]{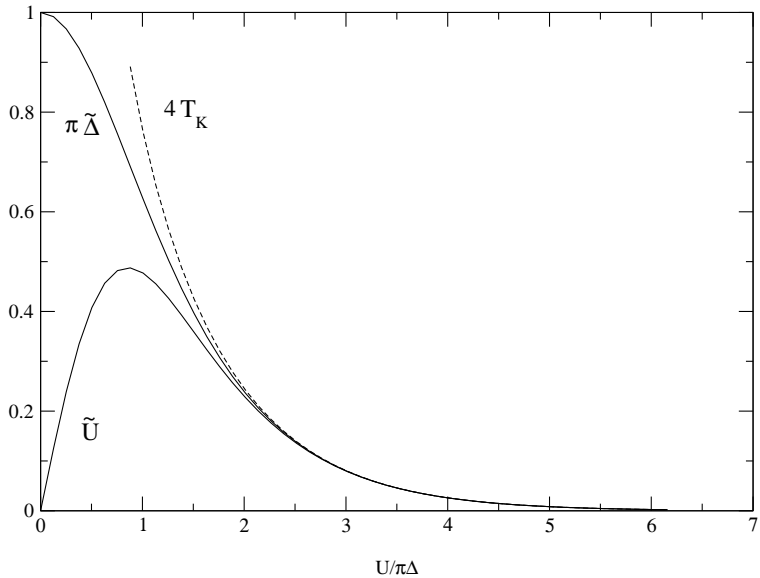}
\end{center}
\caption{A plot of the renormalized parameters $\tilde U$ and $\pi\tilde\Delta$ 
 for the symmetric Anderson model in terms of the bare parameters $U$ 
 and $\pi\Delta$. In the comparison of these parameters with $4T_{\rm K}$
 for  $U\gg \pi\Delta$  the  value of $T_{\rm K}$  is given in
 equation (\ref{tk}).}
\end{figure}

To calculate low temperature conductivity to order $T^2$ one needs to
calculate the renormalized self-energy to order $\omega^2$ 
and $T^2$. There is a $T^2$ contribution to the conductivity arising from
the scattering of free quasiparticles by the impurity  but there is an additional contribution 
 due to the mutual scattering of the quasiparticles due to the 
inter-quasiparticle interactions.   The lowest order contribution of
this type arises from the second order diagram for $\tilde\Sigma$
 shown in figure 2(b). The only counter term diagram that has to be taken
into account to order $\omega^2$ or $T^2$ is due to the second term
in $\lambda_2^{(2)}$,  which is required to cancel the linear in $\omega$ term
arising from the diagram in figure 2(b), and this gives $\lambda_2^{(2)}=3-\pi^2/4$. We calculate this for the case of particle-hole symmetry which
is such that $\tilde\epsilon_d=\tilde U/2$ and $n_d(T)=1$.
The contribution to the imaginary part of the renormalized self-energy
from 2(b) is
\begin{equation}{\rm Im}\tilde\Sigma(\omega,T)=\pi\tilde U^2\int 
\tilde\rho_d(\epsilon)\tilde\rho_d(\epsilon')\tilde\rho_d(\omega-\epsilon-\epsilon')
D(\omega,\epsilon,\epsilon')d\epsilon d\epsilon',\label{2self}\end{equation}

where
\begin{equation}D(\omega,\epsilon,\epsilon')=(1-f(\epsilon)-f(\epsilon'))f(\epsilon+\epsilon'-\omega)+f(\epsilon)f(\epsilon'),\label{ff},\end{equation}
and $f(\epsilon)$ is the Fermi factor $1/(1+e^{\beta\epsilon})$ with $\beta=1/T$.
To calculate the conductivity to order $T^2$ we need to evaluate this integral to order
$\omega^2$ for $T=0$.  For $T=0$ equation (\ref{ff}) becomes
\begin{equation}D(\omega,\epsilon,\epsilon')=(\theta(\epsilon)+\theta(\epsilon'))(\theta(\omega-\epsilon-\epsilon'))+(1-\theta(\epsilon))(1-\theta(\epsilon')),\end{equation}
where $\theta(x)$  is the step function. To find the $\omega^2$ coefficient we differentiate
twice with respect to $\omega$ and use the relations,
\begin{equation}{\partial\theta(x)\over \partial x}=\delta(x),\quad{\partial^2\theta(x)\over \partial x^2}=\delta'(x).\end{equation}
As $\tilde \rho_d(0)=1/\pi\tilde\Delta$ for the case of particle-hole symmetry,  the result is
\begin{equation}{\rm Im}\tilde\Sigma(\omega,0)=-{\pi\over 2}\tilde U^2\tilde\rho_d(0)^3\omega^2
=-{\omega^2\over 2\tilde\Delta}\left({\tilde U\over \pi\tilde\Delta}\right)^2.\label{omega2}\end{equation}
We need the corresponding results to order $T^2$ for $\omega=0$. For $\omega=0$
the temperature dependent factor in the integrand of (\ref{ff}) is
\begin{equation}D(0,\epsilon,\epsilon')=2f(\epsilon)f(\epsilon')(1-f(\epsilon+\epsilon'))
,\end{equation}
 We can change the
variables of integration to $x$ and $x'$, where $x=\epsilon T$ and $x'=\epsilon'   T$, and the
integral of equation (\ref{2self}) to order $T^2$  becomes
\begin{equation} {\rm Im}\tilde\Sigma(0,T)=-{ T^2\over \tilde\Delta}\left({U\over \pi\tilde\Delta}\right)^2\int_{-\infty}^{\infty} \int_{-\infty}^{\infty} 
2F(x)F(x')(1-F(x+x'))dx dx',\label{}\end{equation}
where $F(x)=1/(1+e^x)$. The integration over $x'$ can be carried out to give
\begin{equation} {\rm Im}\tilde\Sigma(0,T)={ T^2\over \tilde\Delta}\left({U\over \pi\tilde\Delta}\right)^2\int_{-\infty}^{\infty} 
{x\over{\rm  sinh}x}dx ={\pi^2 T^2\over 2\tilde\Delta}\left({U\over \pi\tilde\Delta}\right)^2.\end{equation}
To evaluate the conductivity we need to evaluate the transport relaxation life-time
$\tau(\omega,T)$ which is proportional to the inverse of the impurity density of states
$\rho_d(\omega,T)$ which in turn is proportional to the imaginary part of the renormalized
Green's function so that
$$\tau(\omega,T)\propto \pi\tilde\rho_d(\omega,T)^{-1}=\tilde\Delta -{\rm Im}\tilde\Sigma(\omega,T)+{{(\omega-{\rm Re}\tilde\Sigma(\omega,T))^2}\over{(\tilde\Delta -{\rm Im}\tilde\Sigma(\omega,T))}}.$$
\begin{equation}\hfill\end{equation}
When these results  are used
 to evaluate the contribution to the impurity conductivity $\sigma_{\rm imp}(T)$  to order $T^2$ we find   \begin{equation}\sigma_{\rm imp}(T)
=\sigma_0\biggl\{1+{\pi^2\over3}\left({{T}\over{\tilde\Delta}}
\right)^2(1+2(R-1)^2)+{\rm O}(T^4)\biggr\}.\label{(52)}\end{equation}
\smallskip\noindent where $R$ is the Wilson ratio  given $R=1+\tilde U/\pi\tilde\Delta$.
This is an exact result to order $T^2$ which was first derived by Nozi\` eres \cite{noz} for the Kondo regime, which corresponds to $\tilde U=\pi\tilde\Delta=4T_{\rm K}$
and $R=2$.  The more
general  result was derived  by Yamada \cite{yam}. 
More recently Lesage and Saleur \cite{ls} have also calculated the coefficients of the $T^4$ and $T^6$ terms
in this expansion in the Kondo regime, using boundary conformal field theory. \par
Nothing has been omitted in the renormalized perturbation expansion, and it gives the asymptotically exact results in the low temperature regime, when taken to second order in $\tilde U$, so it would be interesting to extend the results by including higher order terms. One
possibility would be to include all the terms to fourth order in $\tilde U$, and calculate the
coefficient of the next correction term in the conductivity, the $T^4$ term, to compare the result with that of Lesage and Saleur. However,
this would be require an expansion of the self-energy in terms of both the frequency and temperature for all
the fourth order terms, which, though straight forward to carry out, would be a rather long and tedious exercise. An alternative way of examining the contributions from the next order terms would be  to calculate the $H^3$ term in field dependence of the impurity magnetization
in a weak magnetic field. The linear term
in $H$ was given exactly by the first order renormalized expansion. The coefficient of the
 $H^3$ term  is known exactly for the Kondo model at $T=0$, and also for the symmetric Anderson model, from Bethe ansatz calculations \cite{ba,tw}. The renormalized perturbation calculation
of this coefficient to order $\tilde U^3$  is described in the next section.\par

\section {\bf Higher order terms}\par

We will perform the renormalized perturbation calculations here in a slightly different but equivalent way from the one used in the previous section. It will have the advantage of not involving  the explicit use of counter-terms. We will
also obtain an  expression for the
renormalized parameters in terms of the bare ones, at least for weak coupling. We first of all use the standard perturbation
theory in $U$, and then calculate the renormalized parameters explicitly to the appropriate order.
We can invert this relation and then write the standard perturbation result in  terms of the renormalized parameters, i.e. we renormalize the standard perturbation terms order by order using the renormalization
conditions (\ref{ren1}) and (\ref{ren2}). The result will correspond to the renormalized expansion in $\tilde U$, as described in the previous section, when taken
to the same order.\par
\subsection{Third order perturbation theory}\par
We use the zero temperature formalism where the impurity Green's function can be written in the form,
   \begin{equation}
G_{\sigma}(\omega,h)={1\over{\omega-\epsilon_{\rm d}+\sigma h+i\Delta{\rm sgn}(\omega)
+\Sigma_{\sigma}(\omega, h)}},\end{equation}
where $h=g\mu_{\rm B}H/2$.
An expression for the impurity magnetization in terms of the magnetic field dependent self-energy
at $T=0$ can be derived from the Friedel sum rule, where the
 impurity level occupation number $n_{{\rm d},\sigma}$ in the spin channel 
$\sigma$   is given by
\begin{equation} n_{{\rm d},\sigma}={1\over
2}-{1\over\pi}\tan ^{-1}\left
({\epsilon_d -\sigma h+\Sigma_{\sigma}(0,h)\over{\Delta}}\right
),\label{fsr}\end{equation}
which is equivalent to equation (\ref{rfsr}).
We can deduce from this an expression for the induced impurity magnetization,
and expand it to order $h^3$. However, it will be useful to separate out
the skeleton tadpole diagram shown in figure 2(a), which has the full Green's function, indicated
by a double propagator, in the bubble, as this is equal to $Un_{d,-\sigma}$,
where $n_{d,-\sigma}$ is the exact expectation value of the occupation number.
We write the self-energy $\Sigma_\sigma(\omega, h)$ in the form,
\begin{equation}
\Sigma_\sigma(\omega, h)=Un_{d,-\sigma}+\bar \Sigma_\sigma(\omega, h),
\end{equation}
  and substitute it into equation (\ref{fsr}). We write
the impurity magnetization $M(h)=g\mu_{\rm B}(n_{d,\uparrow}-n_{d,\downarrow})/2$, in a weak field as a power series,
\begin{equation}
M(h)={g\mu_{\rm B}\over\pi}\sum_n M_{2n+1}\left({h\over \Delta}\right)^{2n+1}.\label{Mn}
\end{equation}
\begin{figure}
\hspace{1.8cm}
\includegraphics{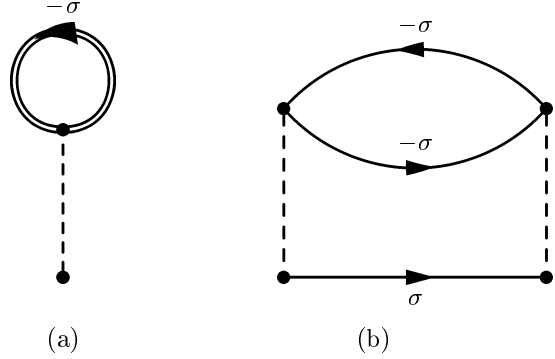}

\caption{ The skeleton tadpole diagram, (a), and second order self-energy diagram, (b).}
\end{figure}
For the particle-hole symmetric model ($\epsilon_d=U/2$) these coefficients
in terms of the self-energy $\bar\Sigma_{\uparrow}(0,h)$ is
\begin{equation}
M_1={1\over{(1-U/\pi\Delta)}}\left.\left(1-{\partial\bar\Sigma_{\uparrow}(0,h)\over
\partial h}\right|_{h=0}\right),\end{equation}
\begin{equation}
M_3=- {1\over{3(1-U/\pi\Delta)}}\left. \left(M_1^3+{\Delta^2\over 2}{\partial^3\bar\Sigma_{\uparrow}(0,h)\over
\partial^3h}\right|_{h=0}\right).\end{equation}
We have the results for the first
derivative of the self-energy with respect to $h$ to order $U^3$ from the calculations of
of Yamada \cite{yam},
\begin{equation}M_1=1+{U\over\pi\Delta}+\left(3-{\pi^2\over 4}\right)\left({U\over\pi\Delta}\right)^2+\left(15-{3\pi^2\over 2}\right) \left({U\over\pi\Delta}\right)^3.
\end{equation}
The only unknown term in the expression for the third order magnetization to  order $U^3$ is the third order derivative of the self-energy at zero frequency with respect to the magnetic field $h$.

To second order in $U$ there is only one diagram which contributes to $\bar\Sigma_{\sigma}(\omega,h)$, that 
shown in figure 2(b), which gives a contribution,
 \begin{equation}\bar\Sigma_{\uparrow}^{(2b)}(\omega,h)=U^2\int G^{(0)}_{\uparrow}(\omega-\omega',h)\Pi^{p\downarrow,h\downarrow}
(\omega',h){d\omega'\over 2\pi i}.\end{equation}
The particle-hole $\Pi^{p\sigma,h\sigma'}$ propagator and the corresponding  particle-particle propagator $\Pi^{p\sigma,h\sigma'}$ are both evaluated in appendix A.
 \begin{equation}\bar\Sigma_{\uparrow}^{(2b)}(0,h)=-h\left(2-{\pi^2\over4}\right)\left({U\over\pi\Delta}\right)^2+
{Ch^3\over{3\Delta^2}}\left({U\over\pi\Delta}\right)^2,\end{equation}
where the coefficient $C$ has been evaluated numerically, and we find a value
$C\approx -1.735$.\par

The third order
diagrams fall into two  types.
 There are three diagrams shown in figure 3
corresponding to dressing each of the propagators in the second order self-energy diagram 
with a simple tadpole or zero order Hartree term.
For the symmetric model the contributions from the first two diagrams, 3(a) and 3(b),
cancel to first order in $h$ but contribute to higher order. The contribution from the diagram 3(a) in which the particle line is
dressed with a tadpole is
\begin{equation}\bar\Sigma_{\uparrow}^{(3a)}(\omega,h)={U^3\over\pi}{\rm tan}^{-1}\left({h\over\Delta}\right)\int (G^{(0)}_{\downarrow}(\omega-\omega',h))^2\Pi^{p\uparrow,h\downarrow}(\omega',h){d\omega'\over 2\pi i}
.\end{equation}
\begin{figure}
\begin{center}
\includegraphics{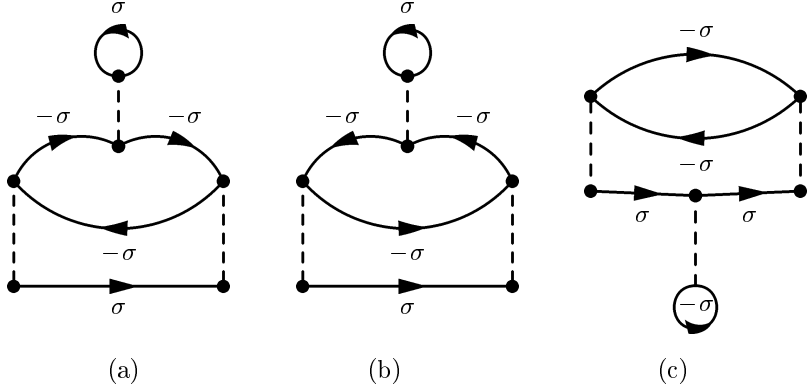}
\end{center}
\caption{Third order diagrams with a tadpole insertion. }
\end{figure}
The contribution from the corresponding diagram 3(b) in which the spin-down hole line is dressed is
\begin{equation}\bar\Sigma_{\uparrow}^{(3b)}(\omega,h)={U^3\over\pi}{\rm tan}^{-1}\left({h\over\Delta}\right)\int( G^{(0)}_{\downarrow}(-\omega+\omega',h))^2\Pi^{p\uparrow,p\downarrow}(\omega',h){d\omega'\over 2\pi i}
.\end{equation}
As $\Pi^{p\uparrow,p\downarrow}(\omega',h)=-\Pi^{p\downarrow,h\downarrow}(\omega',h)$
and $G^{(0)}_{\downarrow}(-\omega+\omega',h)=-G^{(0)}_{\uparrow}(\omega-\omega',-h)$, we can rewrite this contribution as
\begin{equation}\bar\Sigma_{\uparrow}^{(3b)}(\omega,h)=-{U^3\over\pi}{\rm tan}^{-1}\left({h\over\Delta}\right)\int (G^{(0)}_{\downarrow}(\omega-\omega',-h))^2\Pi^{p\downarrow,h\downarrow}(\omega',h){d\omega'\over 2\pi i}
.\end{equation}
We are left with the contribution from the diagram in figure 3(c) in which
the spin $\uparrow$ propagator  of a particle-hole bubble is dressed
with a Hartree bubble.
\begin{equation}\bar\Sigma_{\uparrow}^{(3c)}(\omega,h)=-{U^3\over\pi}{\rm tan}^{-1}\left({h\over\Delta}\right)\int (G^{(0)}_{\uparrow}(\omega-\omega',h))^2\Pi^{p\downarrow,h\downarrow}
(\omega',h){d\omega'\over 2\pi i}.\end{equation}
The total result from the three diagrams to order $h^3$ is
\begin{equation}\bar\Sigma_{\uparrow}^{(3)}(0,h)=-h\left(2-{\pi^2\over4}\right)\left({U\over\pi\Delta}\right)^3
+{Eh^3\over 3\Delta^2}
\left({U\over\pi\Delta}\right)^3,\end{equation} 
 where the coefficient E is calculated numerically as -5.670.\par 
Finally there are the two diagrams illustrated in figure 4(a) and (b), which
 can be regarded as being derived from
the second order diagram, 2(b)  with an intermediate scattering in the subdiagram
corresponding to one of the dynamic susceptibilities. The contribution from the diagram in 
figure 4(a) is
 \begin{equation}\bar\Sigma_{\uparrow}^{(4a)}(\omega,h)=-U^3\int G^{(0)}_{\downarrow}(\omega-\omega',h)
(\Pi^{p\uparrow,h\downarrow}
(\omega',h))^2{d\omega'\over 2\pi i},\end{equation}
with intermediate particle-hole scattering.
 For the diagram in figure 4(b) with intermediate particle-particle scattering the contribution is
\begin{equation}\bar\Sigma_{\uparrow}^{(4b)}(\omega,h)=-U^3\int G^{(0)}_{\downarrow}(
\omega'-\omega,h)(\Pi^{p\uparrow,p\downarrow}
(\omega',h))^2{d\omega'\over 2\pi i}.\end{equation}
The total result is
\begin{equation}\bar\Sigma_{\uparrow}^{(4a)}(0,h)+\bar\Sigma_{\uparrow}^{(4b)}(0,h)=-h(10-\pi^2)\left({U\over\pi\Delta}\right)^3
+{Dh^3\over 3\Delta^2}\left({U\over\pi\Delta}\right)^3.\end{equation} 
The coefficient D was estimated numerically as 
 D=1.541.\par
Collecting the results to third order in $U$ together,
\begin{equation}
M_3=-\left\{1+
4{U\over\pi\Delta}+A\left({U\over\pi\Delta}\right)^2+
 B\left({U\over\pi\Delta}\right)^3
\right\},\end{equation}
\begin{equation}A=16-{3\pi^2\over 4}+C,\quad B=80-{27\pi^2\over 4}+C+D+E.\end{equation}
The coefficients A and B can also be  deduced from the Bethe ansatz 
results for the magnetization of the symmetric Anderson model by generalizing the approach of Horvati\'c and Zlati\'c \cite{hz}, who derived a recurrence relation for a series expansion in powers of $U$ for the coefficient
$M_1$, to obtain a similar expansion for $M_3$. The details are given in  Appendix B.  
 We find $A =65/3-3\pi^2/2$ which gives $C=17/3-3\pi^2/4=-1.7355$ in
complete agreement with the numerical estimate, and 
$B=15(280/27-\pi^2)= 7.512$  which agrees well the numerical estimate 7.515.\par
\subsection{Renormalization}\par
Having calculated all the self-energy terms to order $U^3$ using the standard
perturbation theory we want to deduce the corresponding results in the renormalized expansion to order $\tilde U^3$.
 \begin{figure}
\begin{center}
\includegraphics{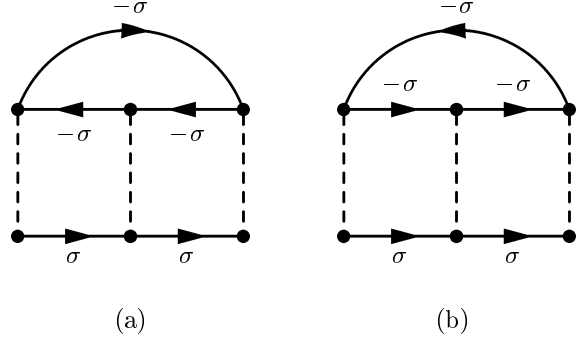}
\end{center}
\caption{Third order diagrams with repeated particle-hole, (a), and particle-particle scattering, (b). }
\end{figure}

We need  to calculate the renormalized parameters, $\tilde\Delta$ and
 $\tilde U$, using  the definitions given in equations (\ref{ren1}) and (\ref{ren2}), in terms of the bare parameters to order  $U^3$. For this we will need the  wavefunction renormalization factor $z$ and the irreducible
four point vertex $\Gamma_{\uparrow,\downarrow}(0,0,0,0)$.
The only contribution to $z$ to  order $U^3$ comes from the
second order diagram and the result  is
\begin{equation}z=1-\left(3-{\pi^2\over 4}\right)\left({U\over\pi\Delta}
\right)^2+{\rm O}\left[\left({U\over\pi\Delta}
\right)^4\right]\end{equation}
which can also be deduced from the results in  Yamada's paper \cite{yam}.
The contributions from the diagrams for the irreducible vertex $\Gamma_{\uparrow,\downarrow}(0,0,0,0)$ 
to second order in $U$ cancel in the absence of a magnetic field. Diagrams which contribute at third order in $U$ are shown in figure 5. There are three possible diagrams of the type shown in figure 5(a)
 each one giving a contribution $U(U/\pi\Delta)^2$. There are six other diagrams in all of the type shown in figure 5(b) and (c), and each of these gives a contribution $(2-\pi^2/4)U(U/\pi\Delta)^2$. The total to third order in $U$
is 
\begin{equation}\Gamma_{\uparrow,\downarrow}(0,0,0,0)=U\left\{1+\left(15-{3\pi^2\over 2}
\right)\left({U\over\pi\Delta}\right)^2....\right\}.\end{equation}
From these two results we can deduce the renormalized parameters to order $U^3$,
\begin{equation}\tilde\Delta=\Delta\left\{1-\left(3-{\pi^2\over 4}\right)\left({U\over\pi\Delta}
\right)^2....\right\},\quad\tilde U=U\left\{1-(\pi^2-9)\left({U\over\pi\Delta}\right)^2...
\right\}.\end{equation}
These results correspond to the weak coupling region $U\ll \pi\Delta$
in the plot of the renormalized parameters shown in figure 1.\par
We can invert these expression to deduce the bare parameters $\Delta$ and $U$ in terms of
the renormalized ones, $\tilde\Delta$ and $\tilde U$,
\begin{equation}\Delta=\tilde\Delta\left\{1+\left(3-{\pi^2\over 4}
\right)\left({\tilde U\over\pi\tilde\Delta}
\right)^2....\right\},\quad U=\tilde U\left\{1+(\pi^2-9)\left({\tilde U\over
\pi\tilde\Delta}\right)^2...
\right\}.\end{equation}
\begin{figure}
\begin{center}
\begin{tabular}[t]{cc}
\includegraphics{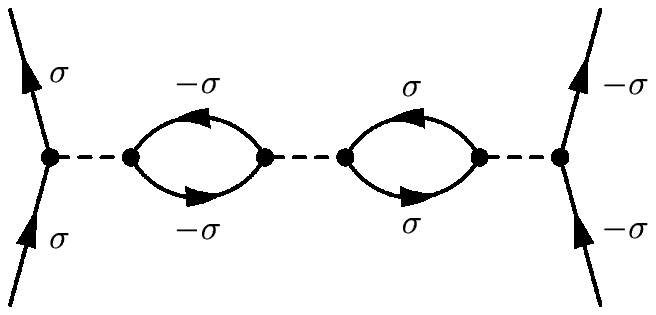}&
\includegraphics{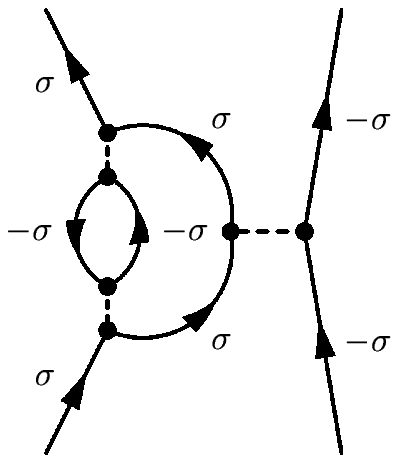}\\
(a) & (b) 
\end{tabular}
\includegraphics{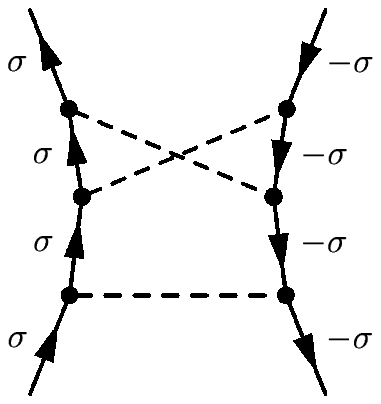} \\
(c)
\end{center}
\caption{Third order contributions to the irreducible four-vertex $\Gamma_{\uparrow,\downarrow}(0,0,0,0)$.  }
\end{figure}
We now use these relations to express the third order result in $U$ for the magnetization 
in terms of the renormalized parameters. It will be convenient to write (\ref{Mn}) in a modified
form
\begin{equation}
M(h)={g\mu_{\rm B}\over\pi}\sum_n\tilde M_{2n+1}\left({h\over\tilde \Delta}\right)^{2n+1}.
\end{equation}
The coefficient $\bar M_3$ to third order in
$\tilde U$ is given by
\begin{equation}
\tilde M_3=
\left\{1+4\left({\tilde U\over\pi\tilde\Delta}
\right)
+A'\left({\tilde U\over\pi\tilde\Delta}\right)^2+
B'\left({\tilde U\over\pi\tilde\Delta}\right)^3
\right\},\end{equation}
where
\begin{equation}A'=7+C,\quad B'={5\pi^2\over 4}-4+C+D+E.
\end{equation}
We know from the Ward identity (\ref{wi}) that the result to  order $h$ is exact to all orders in $\tilde U$. We know that  the term of order $h^3$ ia asymptotically exact in the weak coupling regime, $U/\pi\Delta\to 0$, so it is of interest to check it in the strong coupling limit against exact Bethe ansatz results for the Kondo model. We use the results in the previous section to express all the renormalized parameters in terms of the Kondo temperature $T_{\rm K}$
  $\tilde U/\pi\tilde\Delta\to 1$ and $\pi\tilde\Delta\to 4T_{\rm K}$ as $U\to \infty$.
We then find 
\begin{equation}{ M(h)\over g\mu_{\rm B}}={h\over 2T_{\rm K}}- D'{h^3\pi^2\over  192T_{\rm K}^3},
\label{kon}\end{equation}
where $D'=5+A'+B'=8+{5\pi^2/4}+2C+D+E$.
 With the values of the coefficients as deduced from the third order renormalized perturbation theory we get $D'=12.73$. If the exact Bethe ansatz result is written in the same form, using the same definition of
$T_{\rm K}$, the coefficient $D'$ has the value $D'=24\sqrt{3}/\pi=13.232$. The  error from our third order results is less than 4\%. Hence the perturbation theory taken to third order, which is asymptotically exact in the weak coupling regime, is
 very close to the exact result at strong coupling. \par
As $\tilde U/\pi\tilde\Delta\to 1$ in the strong coupling limit, the factor  $(\tilde U/\pi\tilde\Delta)^n$ multiplying the contributions from the nth order terms
do not decrease with $n$, as they do in the weak coupling limit $\tilde U/\pi\tilde\Delta\ll 1$. In the Appendix B we show that in the
Kondo limit no finite set of renormalized diagrams can give the 
$h^3$ coefficient in the magnetization exactly. However, we have shown that the error is small in the limit when the perturbation series is taken to third order, and is even smaller  for intermediate and weak coupling. This is clearly seen from the results in figure 6 where we plot the coefficient $\tilde M_3$ against $\tilde U/\pi\tilde\Delta$ over the range from weak ($\tilde U/\pi\tilde\Delta<<1$) to strong coupling ($\tilde U/\pi\tilde\Delta=1$) as calculated from the third
order renormalized perturbation theory and compare it with the exact Bethe ansatz results, expressed in terms of the renormalized parameters $\tilde U$ and $\tilde\Delta$. Over the range from $U=\tilde U=0$
to $U=5\pi\Delta$, ($0<\tilde U/\pi\tilde\Delta\le 0.9998$), $\tilde\Delta/\Delta$
varies by two orders of magnitude, from 1 to $\sim 8\times 10^{-3}$. \par
\begin{figure}
\begin{center}
\includegraphics[width=10cm]{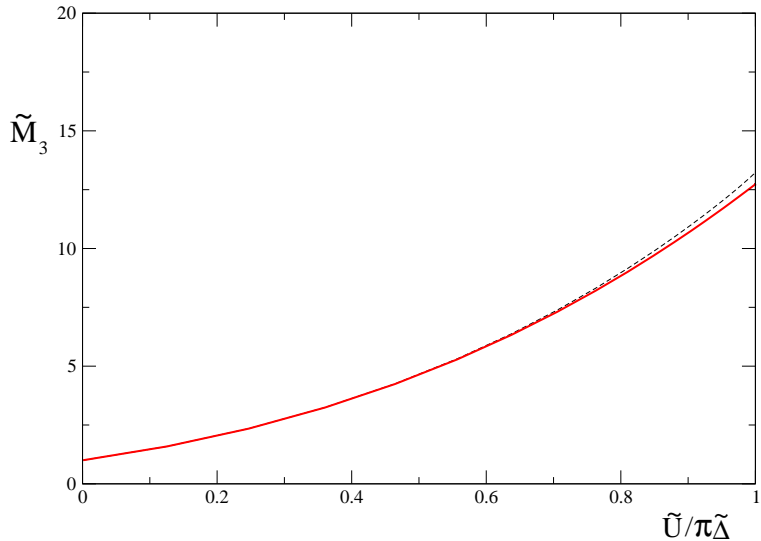}
\end{center}
\caption{The coefficient $\tilde M_3$ from the third order renormalized calculation
compared with the exact Bethe ansatz result (dotted line) plotted as a function of $\tilde U/\pi\tilde\Delta$. }
\end{figure}
It would be
interesting if there would be some way of extracting the small $h^3$ correction from the higher order diagrams. We have taken account of the contribution from the leading order irrelevant term ($\tilde U$) about the low energy fixed point to third order, so that the remaining contributions  must be related to the next order irrelevant terms in the effective Hamiltonian at the Wilson strong coupling fixed point. These terms must be a combination
of local operators but it is not clear how to relate these explicitly to the higher order
diagrams in the renormalized perturbation theory.\par

\section{Dynamic response functions in a weak  magnetic field}
\par
If we calculate the $\omega$-dependence of the self-energy as well as the
h-dependence we can deduce the form of the quasiparticle density of states
in a weak magnetic field. If we calculate this as a general function of 
$\omega$ and $h$, rather than expanding in powers of $\omega$ and $h$, it will be more
convenient to revert to  the renormalized expansion as used in section 2 with the explicit use of counter-terms. This just requires a rearrangement of the terms
calculated in the previous section. Each diagram will now be interpreted as a diagram for the renormalized
self-energy $\tilde\Sigma_{\sigma}(\omega, h)$, with $U\to \tilde U$ and
$\Delta\to \tilde\Delta$, and $\tilde\epsilon_d=-\tilde U/2$ but we will have to include the counter-terms to order
$\tilde U^3$ to satisfy the renormalization conditions. For the particle-hole symmetric model, the only non-zero 
counter-terms to third order are:  $\lambda_1=0$, $\lambda_2^{(2)}=(3-\pi^2/4)$,
and $\lambda_3^{(3)}=-\pi\tilde\Delta(15-3\pi^2/2)$. The only new term to order $\tilde U^3$ is the last term which cancels off the renormalization 
of the the four vertex $\tilde\Gamma_{\uparrow,\downarrow}(0,0,0,0)$ shown in figure 5, which is
not needed as the vertex $\tilde U$ is taken to be the fully renormalized one. An alternative way to calculate the counter-terms is directly from their definitions  in terms of the self-energy and vertex functions, and to re-express these in term of the renormalized parameters. \par
The counter term diagrams are shown in figure 7. The first diagram 7(a) involves the $\lambda_2$
vertex, and ensures that the linear term in $\omega$ is cancelled off. The next diagram 7(b) is an
additional tadpole contribution arising from the counter-term interaction $\lambda_3$. There is  also a third order counter-term diagram 7(c) arising from a combination of the tadpole diagram to
order $\tilde U$ with a counter-term vertex $\lambda^{2}$ on the bubble. This diagram gives a
contribution
\begin{equation}\Sigma_{\uparrow}^{\rm ct}(\omega,h)=(3-{\pi^2\over 4})\tilde U\left({\tilde U\over \pi\tilde\Delta}\right)^2\int (\omega'+h)(G^{(0)}_{\downarrow}(
\omega',h))^2{d\omega'\over 2\pi i}.\end{equation}
The evaluation of the integral is straight forward and gives
\begin{equation}
\Sigma_{\uparrow}^{\rm ct}(\omega,h)=\tilde\Delta(3-{\pi^2\over 4})\left({\tilde U\over \pi\tilde\Delta}\right)^3\left[{\rm tan}^{-1}\left({h\over\tilde\Delta}\right)-{\tilde\Delta h\over {h^2 +\tilde\Delta^2}}\right].
\end{equation}
\begin{figure}
\hspace{1cm}
\includegraphics{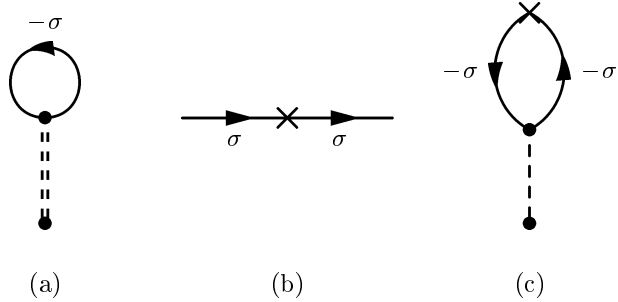} 
\caption{Counter-term diagrams which contribute to the self-energy  to third order
in $\tilde U$. The double-dashed line represents the vertex $\lambda_3$,  the cross represents the vertex $\lambda_2$, and the single dashed line $\tilde U$.}
\end{figure}
 If the self-energy to third order from the standard diagrams
calculated in the previous section in terms of the renormalized parameters is written as $\Sigma^{(3)}_{\uparrow}(\omega, h)$, then the renormalized self-energy to order $\tilde U^3$ is
given by
$$\tilde\Sigma_{\uparrow}^{(3)}(\omega,h)=\Sigma_{\uparrow}^{(3)}(\omega,h)+\tilde\Delta(15-{3\pi^2\over 2}){\rm tan}^{-1}\left({h\over\tilde\Delta}\right)\left({\tilde U\over \pi\tilde\Delta}\right)^3 
$$
\begin{equation}+(\omega+h)(3-{\pi^2\over 4})\left({\tilde U\over \pi\tilde\Delta}\right)^2 
-\tilde\Delta(3-{\pi^2\over 4})\left[{\rm tan}^{-1}\left({h\over\tilde\Delta}\right)-{\tilde\Delta h\over {h^2 +\tilde\Delta^2}}\right]\left({\tilde U\over \pi\tilde\Delta}\right)^3.
\end{equation}
One can check that, in the limit $\omega=0$ and expanded to order $h^3$, this renormalized self-energy,
when substituted into equation (\ref{rfsr}), gives the same results for the
magnetization as were obtained in the previous section. \par
In figure 8 we plot the quasiparticle spectral densities in weak and strong
coupling as a function of $\omega/\tilde\Delta$ for $h=0.15\tilde\Delta$. The peaks in the spectral density shift from $\omega_{\rm max}=\pm h$ at weak coupling to 
$\omega_{\rm max}=\pm 4h/3$, at strong coupling. These are asymptotically exact results as $h\to 0$, as
has been shown by Logan and Dickens \cite{ld}. The general result for the position 
of the maximum in weak field from their calculation can be written in the form,
\begin{equation}\omega_{\rm max}={\pm 2h \left(1+{\tilde U\over\pi\tilde\Delta}\right)\over{2+\left(\tilde U\over\pi\tilde\Delta\right)^2}}.\end{equation}
The term in $\tilde U^2$ in the denominator arises from the contribution from the imaginary
part of the self-energy at low frequency.
The peak in the spectral density in the strong coupling $\tilde U/\pi\tilde\Delta=1$, or localized limit, is the Kondo resonance, which has a width, $\tilde\Delta=4 T_{\rm K}/\pi$.\par
\begin{figure}
\begin{center}
\includegraphics[width=10cm]{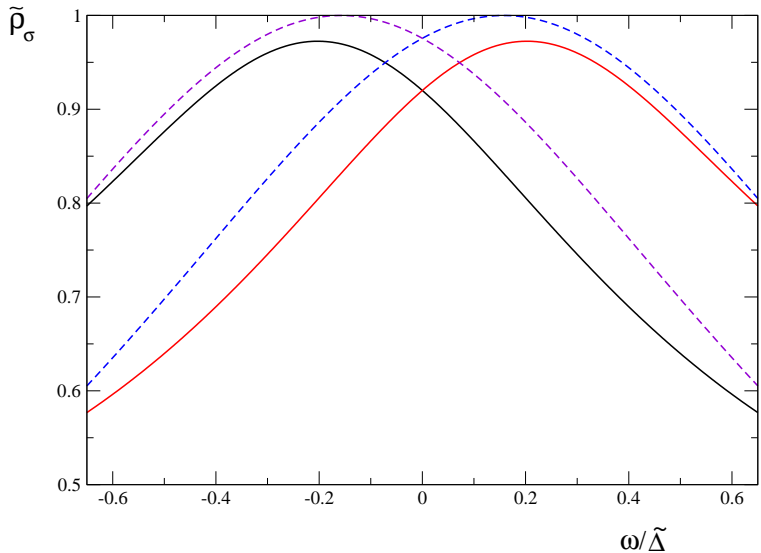}
\end{center}
\caption{ The density of states f0r $\uparrow$ and $\downarrow$ quasiparticles in a magnetic field $h/\tilde\Delta=0.15$ for weak coupling $\tilde U/\pi\tilde\Delta=0.1$ (dotted curves) and strong coupling $\tilde U/\pi\tilde\Delta=1.0$ plotted as a function of $\omega/\tilde\Delta$. }
\end{figure}
There have been a number of recent calculations of the Kondo resonance in a magnetic field
for the Anderson model. The surge of interest in this topic has been due to the recent observations of the Kondo effect in quantum dots \cite{ggk,fran}. These mesoscopic systems can be described
by an impurity Anderson model coupled by leads to two electron reservoirs, and  tunnelling through these dots at very low temperatures is possible due to the presence of the Kondo resonance. As the parameters
in the Anderson model description of a quantum dot depend upon gate voltages, they can be
modified is a much more controlled way than for a real magnetic impurities, which should present
a greater range of possibilities for  comparing  theory with experiment. There have been been
numerical
renormalization group calculations
of the Kondo resonance in a magnetic field  \cite{cos,hof}, approximate treatments based on  the Bethe ansatz equations \cite{mw,ksl},
and also results from the local moment approach \cite{lma}. The advantage of the renormalized perturbation approach is
that the results are asymptotically exact in the limits of small $\omega$ and $H$, and
form a useful check on the other methods.\par

\section{Conclusions}\par
We have shown that the renormalized perturbation theory (RPT) approach provides a way of going beyond the normal limitations of standard  perturbation theory. The potential of this
approach has been illustrated in the particular case of the Anderson impurity model,
where we have shown that low order RPT calculations  provide
a comprehensive description of the low energy, low temperature range, in the Kondo as well as the 
weak coupling regime. Except in the weak coupling regime, we do not have explicit expressions
for the renormalized parameters in terms of the bare ones (other than than deducing them from the Bethe ansatz results, see figure 1), but this is also true of other
approaches to magnetic impurity problems such as the conformal field theory \cite{ls}. There
may be ways of estimating the renormalized parameters using variational methods, numerically, or by the summing a subset of diagrams, as in the local moment approach which gives a good approximate interpolation from weak to strong coupling.  The RPT approach does give a clear physical
picture of the Fermi-liquid regime, and the low order results are asymptotically exact in this limit. This is also the case for other magnetic impurity models that have been studied, which include
degeneracy and extra interactions, such as a Hund's rule coupling, and explicit expressions
have been derived for the renormalized interactions in terms of the Kondo temperature, in the
strong coupling limit \cite{h-rpt,hew}. It provides a complementary approach to the Wilson style of calculations which involve the explicit elimination of higher order excitations \cite{sak,costi}. \par
Two obvious questions arise: Can it be applied to lattice models? 
Is it applicable to systems with a non-Fermi-liquid fixed point?
 The method has also been already been extended to translationally invariant
systems \cite{h-rg}, and related to Fermi-liquid theory. It is an alternative to the Wilson style of renormalization group approach, which has been applied to translationally invariant systems at one loop level by Shankar \cite{sha}. There is potential for applications
here, using the technique described in section 3, for applying the RPT approach to the strong coupling regime for such models
as the Hubbard and periodic Anderson model. The complications that arise for translationally
invariant systems and lattice models are mainly due to the dependence of the self-energy and renormalized vertices on the wave vector ${\bf k}$. There should be some simplification that one could exploit for infinite dimensional models where the self-energy is ${\bf k}$-independent,
and the ${\bf k}$-dependence is suppressed at some types of vertices.\par
The assumption of a finite wave-function renormalization factor $z$ in the derivation of
the renormalized expansion restricts our treatment to Fermi-liquids. Deviations from Fermi-liquid
behaviour can have various causes and each case has to be considered on its own merits.
It has proved possible to generalize the approach to a spinless Luttinger liquid \cite{h-rg},
and to the O(3) symmetric Anderson model, which has a marginal Fermi-liquid fixed point \cite{o3}.
\par\bigskip
\noindent{\bf Acknowledgment}\par
\bigskip
I wish to thank 
 to the  EPSRC  for the support of a research grant (GR/J85349), and the Newton Insititute, Cambridge,
where this work was initiated during their six-month programme on `Strongly Correlated Electron Systems'. 
\par

\section{Appendix A}\par
There are two types of sub-diagrams which correspond to zero order spin  susceptibilities,
with propagators defined by 
\begin{equation}
\Pi^{p\sigma,h\sigma'}(\omega,h)=\int_{-\infty}^{\infty}G^{(0)}_{\sigma}(\omega+\omega',h)
G^{(0)}_{\sigma'}(\omega',h){d\omega'\over {2\pi i}},\end{equation}
and
\begin{equation}
\Pi^{p\sigma,p\sigma'}(\omega,h)=\int_{-\infty}^{\infty}G^{(0)}_{\sigma}(\omega-\omega',h)
G^{(0)}_{\sigma'}(\omega',h){d\omega'\over {2\pi i}}.\end{equation}
These integrals can be evaluated analytically and the results are 
$$\Pi^{p\uparrow,h\uparrow}(\omega,h)={-\Delta\over{\pi(h^2+\Delta^2)}}\quad{\rm for}\quad\omega= 0,$$
\begin{equation}={\Delta\over{\pi\omega(\omega+2i\Delta)}}\left\{{\rm ln}\left({{\omega+i\Delta-h}\over{i\Delta-h}}\right)+
{\rm ln}\left({{\omega+i\Delta+h}\over{i\Delta+h}}\right)\right\}\quad{\rm for}\quad\omega\ne 0.\end{equation}
and
$$\Pi^{p\uparrow,h\downarrow}(\omega,h)={1\over 2\pi\Delta}{\rm ln}\left({{i\Delta-h}\over{i\Delta+h}}\right)-{i\over{\pi(i\Delta-h)}}\quad{\rm for}\quad\omega= -2h,$$
\begin{equation}=i{\Delta\over{\pi}}\left\{{1\over{\omega+2h+2i\Delta}}{\rm ln}\left({{\omega+i\Delta+h}\over{i\Delta+h}}\right)-{1\over{\omega+2h}}
{\rm ln}\left({{\omega+i\Delta+h}\over{i\Delta-h}}\right)\right\}\quad{\rm for}\quad\omega\ne -2h.\end{equation}

\noindent For particle-hole symmetry, we have $
\Pi^{p\uparrow,p\downarrow}(\omega,h)=-\Pi^{p\downarrow,h\downarrow}(\omega,h)$, and
$\Pi^{p\downarrow,h\downarrow}(\omega,h)=\Pi^{p\uparrow,h\uparrow}(\omega,h)=\Pi^{p\downarrow,h\downarrow}(\omega,-h)$.

\section{Appendix B}\par
In this appendix we take the expression for the magnetization from the exact Bethe ansatz results
of Tsvelik and Wiegmann \cite{tw} for the symmetric Anderson model, and deduce a power series in $U$ for the 
coefficient of the term in $H^3$,  along the same lines at that originally used by Horvati\'c and Zlati\'c \cite{hz} for the order $H$ term. The result for the magnetization to order $h^3$ can be written in the form,

\begin{equation} {\pi\Delta M(h)\over g\mu_{\rm B}h}=\left(1-{h^2\over 2u\Delta^2}\right)e^{\pi^2 u/8}
J_1(u)+{h^2\over 2u\Delta^2}e^{3\pi^2 u/8}J_3(u),\end{equation}
where $u=U/\pi\Delta$ where
\begin{equation} J_m(u)=\sqrt{2m\over\pi u}\int_0^{\infty}e^{-mx^2/2u}{{\rm cos}({m\pi x\over 2})
\over 1-x^2}dx. \end{equation}

Horvati\'c and Zlati\'c have developed a power series in $u$ for $J_1(u)$,
\begin{equation}J_1(u)=\sum_{n=0}C_nu^n,\quad{\rm where}\quad C_n=(2n-1)C_{n-1}-{\pi^2\over 4}C_{n-2},\end{equation}
with $C_0=C_1=1$, and $C_2=3-{\pi^2/4}$. The other coefficients  to order $u^5$ are
\begin{equation} C_3=15 -{3\pi^2\over 4},\quad C_4=105-
{45\pi^2\over 4}+{\pi^4\over 16},\quad C_5=15(63-7\pi^2+{\pi^4\over 16}).\end{equation}
We can develop an expansion for $J_3(u)$ in a similar way,
\begin{equation}J_3(u)=\sum_{n=0}\bar C_nu^n\quad{\rm  where}\quad\bar C_n={(2n-1)\over 3}\bar C_{n-1}-{\pi^2\over 4}\bar C_{n-2},\end{equation}
with $\bar C_0=1$, $\bar C_1=1/3$,  $\bar C_2={1/3}-{\pi^2/4}$, and
\begin{equation}\bar C_3={5\over 9} -{\pi^2\over 2},
\quad\bar C_4={35\over 27}-
{5\pi^2\over 4}+{\pi^4\over 16},\quad \bar C_5=5\left({7\over 9}-{7\pi^2\over 9}+{\pi^4\over 16}\right).\end{equation}
We can then write the expression for the magnetization to order $h^3$ in the form
\begin{equation} {\pi\Delta M(h)\over g\mu_{\rm B}h}=\sum_{n=0}C_nu^n-{h^2\over 3\Delta^2}\sum_{n=0}A_nu^n,\end{equation}
where $A_n=3(C_{n+1}-\bar C_{n+1})/2$ for $n\ge 0$. The coefficients of the terms in the second series to order $u^4$ are $A_0=1$, $A_1=4$, and
 \begin{equation} A_2={65\over 3}-{3\pi^2\over 2},\quad A_3=15\left({280\over 27} -{\pi^2}\right),\quad A_4=15\left({847\over 9}-{91\pi^2\over 9}+{\pi^4\over 16}\right).\end{equation}
In the Kondo limit  the term proportional to $J_1(u)$ does not contributes to the $h^3$ coefficient, and it can be shown that the asymptotic contribution from the $J_3(u)$ term agrees with the result for the Kondo model (\ref{kon}) with $T_{\rm K}$ defined
by 
\begin{equation}T_{\rm K}=\Delta\sqrt{\pi u\over 2}e^{-\pi^2/8u+1/2u}\label{tk}\end{equation}\par
We can also use these results to deduce from these results the terms in the renormalized
perturbation calculations to higher orders. We will use this approach to find the fourth order
correction to our third order result. We can deduce $\tilde \Delta$ and $\tilde U$, to fourth order in $U$ from the Bethe ansatz results for $\gamma$ and $\chi$. These can be inverted
to calculate the bare parameters $U$ and $\Delta$ in terms of the renormalized ones to the same order in $\tilde U$. The results are
\begin{equation}{1\over \Delta}={1\over \tilde\Delta}\left(1-\left(3-{\pi^2\over 4}\right)\tilde u^2-\left(24+{15\pi^2\over 4}-{5\pi^4\over 8}\right)\tilde u^4...\right)\end{equation}
 \begin{equation}u=\tilde u\left(1-\left(12-{5\pi^2\over 4}\right)\tilde u^3+{\rm O}(\tilde u^5)\right).\end{equation}
We can then use the results above for the $h^3$ term magnetization to fourth
order in $U$, and rewrite them in terms of the renormalized parameters. 
In this way we calculate that the correction from the fourth order terms to  $D'$,  the $H^3$ coefficient in the Kondo limit, is -0.24145, which is 10\% of the third 
 order contribution. As the coefficient in the Kondo regime has a factor $\sqrt{3}$, the exact
result cannot be obtained within any finite order renormalized perturbation calculation,
as results to  finite order in $\tilde U$ can be expressed as rational functions of the coefficients $C_n$ and $A_n$, and these in turn are  rational numbers and powers of $\pi$, which cannot generate to finite order the
irrational number $\sqrt{3}$.\par

\end{document}